%% file: Gaussian.tex
\documentclass[prx,aps,twocolumn,nolongbibliography,superscriptaddress]{revtex4-1}

\usepackage[colorlinks,bookmarks=true,citecolor=blue,linkcolor=red,urlcolor=blue,citecolor=blue]{hyperref}

\usepackage{float}
\usepackage{amsmath}
\usepackage{cancel}
\usepackage{amssymb}
\usepackage{amsfonts}
\usepackage{bbm}
\usepackage{braket}
\usepackage{xcolor}
\usepackage{appendix}
\usepackage{enumitem}
\usepackage{graphicx}
\usepackage{float}
\usepackage{tikz}
\usepackage{algorithm}
\usepackage{algpseudocode}
\usepackage{bbm}
\usepackage{subfigure}
\usepackage{color}
\usepackage{nicematrix}

\begin{document}

\title{Efficient conversion from fermionic Gaussian states to matrix product states}

\author{Tong Liu}
\affiliation{Institute of Physics, Chinese Academy of Sciences, Beijing 100190, China}
\affiliation{School of Physical Sciences, University of Chinese Academy of Sciences, Beijing 100049, China}

\author{Ying-Hai Wu}
\affiliation{School of Physics and Wuhan National High Magnetic Field Center, Huazhong University of Science and Technology, Wuhan 430074, China}

\author{Hong-Hao Tu}
\email{h.tu@lmu.de}
\affiliation{Faculty of Physics and Arnold Sommerfeld Center for Theoretical Physics, Ludwig-Maximilians-Universit\"at M\"unchen, 80333 Munich, Germany}

\author{Tao Xiang}
\email{txiang@iphy.ac.cn}
\affiliation{Institute of Physics, Chinese Academy of Sciences, Beijing 100190, China}
\affiliation{Beijing Academy of Quantum Information Sciences, Beijing 100190, China}
\affiliation{School of Physical Sciences, University of Chinese Academy of Sciences, Beijing 100049, China}

\begin{abstract}
Fermionic Gaussian states are eigenstates of quadratic Hamiltonians and widely used in quantum many-body problems. We propose a highly efficient algorithm that converts fermionic Gaussian states to matrix product states. It can be formulated for finite-size systems without translation invariance, but becomes particularly appealing when applied to infinite systems with translation invariance. If the ground states of a topologically ordered system on infinite cylinders are expressed as matrix product states, then the fixed points of the transfer matrix can be harnessed to filter out the anyon eigenbasis, also known as minimally entangled states. This allows for efficient computation of universal properties, such as entanglement spectrum and modular matrices. The potential of our method is demonstrated by numerical calculations in two chiral spin liquids that have the same topological orders as the bosonic Laughlin and Moore-Read states, respectively. The anyon eigenbasis for the first one has been worked out before and serves as a useful benchmark. The anyon eigenbasis of the second one is, however, not transparent and its successful construction provides a nontrivial corroboration of our method. 
\end{abstract}

\maketitle

\section{Introduction}
\label{sec:introduction}

At the fundamental level, particles in a quantum many-body system interact with each other. However, the essential physics in many cases are captured by a free particle description, notably in the framework of mean-field theory. A plethora of fascinating states can be generated when free fermions roam in diverse lattice potentials and repel each other solely according to Pauli's exclusion principle. In general, the Hamiltonians of free fermions are quadratic in terms of the creation and annihilation operators so they can be diagonalized easily using single-particle basis transformations. The solutions can be expressed as fermionic Gaussian states (FGSs) whose correlations obey the Wick's theorem~\cite{Ring-book,Bach1994,Bravyi2005,Kraus2010a}. If strong correlations invalidate mean-field treatment, many-body problems become very difficult due to exponential growth of the Hilbert space.

In a variety of strongly correlated phases, mean-field theory can still provide useful insight when it is applied on fictitious partons that emerge from breaking each physical object into multiple pieces. FGSs are still useful when the partons are fermionic and described by quadratic Hamiltonians~\cite{Gros1989}. It is necessary to impose physical constraint using Gutzwiller projection such that partons are glued together to become physical objects. While FGSs of partons are simple, the final wave functions are generally not easy to analyze. Another fruitful approach for strongly correlated systems is tensor networks that include matrix product states (MPSs) and projected entangled pair states (PEPSs)~\cite{XiangT-Book,Cirac2021,Verstraete2008,Cirac2009,Schollwoeck2011,Orus2014}. The two methods have their respective strengths and weaknesses but both can provide successful variational Ans\"atze. Due to the apparent differences, these approaches have been almost independent from each other when they were developed and applied in concrete problems.

It is natural to explore if FGSs and tensor networks can be combined to harness their advantages in collaboration. One attempt is to design hybrid ansatz that contains both of them as ingredients~\cite{ChouCP2012,ZhaoHH2017}. Another route is converting (physically motivated or post-optimized) FGSs into tensor networks that may serve as initial states for numerical simulations to facilitate better convergence. Along this line of thought, several methods have been developed to convert FGSs into MPSs~\cite{Fishman2015,WuYH2020,JinHK2020,Aghaei2020,Petrica2021,Nuesseler2021,JinHK2022a} and they are indeed useful in density matrix renormalization group (DMRG) calculations~\cite{White1992,JinHK2021,ChenJY2021,JinHK2022a,JinHK2023,SunRY2024}. In particular, feeding a DMRG program with suitable initial states could simplify the search for the degenerate ground states in a two-dimensional (2D) system with intrinsic topological order~\cite{JinHK2021,ChenJY2021,SunRY2024}. Along the same line of thoughts, the PEPS representations of FGSs have also been investigated~\cite{Kraus2010b,Wahl2013,Dubail2015,Wahl2014,Hackenbroich2020,Mortier2022,Mortier2023,LiJW2023,YangQ2023,Emonts2023a,Emonts2023b,HeY2024}. It is expected that Gaussian fermionic PEPSs will also find applications in optimization algorithms~\cite{JiangHC2008,Jordan2008,Murg2009,Lubasch2014,Corboz2016,Vanderstraeten2016,LiaoHJ2019,Scheb2023}.

In this work, we develop an algorithm that efficiently converts FGS to MPS by making full usage of the FGS correlation matrix formalism (some papers use the term ``covariance matrix'' instead of ``correlation matrix''). While this method is capable of dealing with cases without translation symmetry, its potential can be most clearly demonstrated using translationally invariant infinite matrix product states (iMPSs). It can be applied when each unit cell contains multiple sites such as 2D systems on infinite cylinders. We have achieved significant reduction of computational cost and constructed reliable iMPSs for very large systems. This method also leads to another important discovery about spontaneous symmetry breaking or topological order. It is well-known that the ground states of such systems are degenerate and their MPSs are ``non-injective" in the sense that local matrices generally have more than one blocks in the canonical form~\cite{Perez-Garcia2007}. Nevertheless, injective MPSs can be obtained from non-injective ones by projecting onto each irreducible block in the virtual space. This progress enriches our toolbox for characterizing topological order in Guzwiller pojected FGSs. For a given iMPS representation, ground-state degeneracy manifests itself in the multiplicity of the leading eigenvectors (``fixed points'') of the iMPS transfer matrix. If we perform the projection to construct an injective MPS, its bipartite entanglement entropy would be minimized. For 2D topological order, injective MPSs with this property are called anyon eigenstates (or ``minimally entangled states''~\cite{Keski1993,ZhangY2012}), which are particularly useful for extracting topological data~\cite{Cincio2013,TuHH2013a,Zaletel2013}. As benchmark examples, this procedure is employed to find the anyon eigenbasis in two chiral spin liquid models and characterize their topological orders. The first system has Abelian SU(2)$_1$ topological order and its anyon eigenbasis has been constructed before. The second system has non-Abelian SU(2)$_2$ topological order but its anyon eigenbasis is unknown and clearly illustrates the utility of our approach. These successful examples demonstrate that the complicated task of filtrating anyon eigenbasis of Gutzwiller projected fermionic wave functions with topological order can be automated. In combination with DMRG, this approach can play an important role in studying the topological order of microscopic models.

The rest of this paper is organized as follows. In Sec.~\ref{sec:methods}, we describe the algorithm in detail. It has three main steps: (i) Get the correlation matrix of the FGS to be converted; (ii) Decompose the correlation matrix of the whole system into local projectors and virtual bonds; (iii) Perform mode decimation and reconfigure local projectors and virtual bonds as local tensors. In Sec.~\ref{sec:results}, the algorithm is applied on two chiral spin liquid models to construct their anyon eigenbases from which entanglement spectra are computed. This paper is summarized with an outlook in Sec.~\ref{sec:summary}.

\section{Method}
\label{sec:methods}

This section presents our algorithm in detail. It starts with a fermionic Gaussian state and decomposes it as a series of local Gaussian states. The resulting state is known as a Gaussian MPS~\cite{Schuch2019}, in which local tensors are represented by their correlation matrices. In principle, these correlation matrices can be converted into superpositions of local Fock states (whose coefficients represent local tensors of MPS) faithfully without any truncation or approximation. However, as a Gutzwiller projection will be applied subsequently, which ruins the Gaussian nature of the MPS, truncations will be introduced when the (Gutzwiller projected) local Gaussian states are converted into MPS tensors.

We first introduce FGSs and the correlation matrix formalism in Sec.~\ref{subsec:methodA}. Next we derive and solve the recursion equation for the correlation matrix of the MPS local tensor in Sec.~\ref{subsec:methodB}. Then, we discuss how to generate local tensors from the local correlation matrix and perform truncation in Sec.~\ref{subsec:methodC}. Finally, we propose the mode decimation scheme to accelerate the computation of the MPS local tensor in Sec.~\ref{subsec:methodD} and discuss how to implement the anyon eigenbasis filtration using iMPS in Sec.~\ref{subsec:methodE}.

\subsection{Fermionic Gaussian states and correlation matrix formalism}
\label{subsec:methodA}

Since we will extensively use FGSs in this work, we first give a brief review of the formulation and introduce the necessary concepts. For notational simplicity, we restrict ourselves in Sec.~\ref{sec:methods} to particle-number-conserving FGSs [termed as U(1)-FGSs]. The extension to the more general case with pairing is straightforward but requires some adaptions of the formalism (see, e.g., Ref.~\cite{JinHK2022a}).

We start with a fermionic lattice system. The fermionic creation and annihilation operators are written as $c^{\dagger}_{j}$ and $c_{j}$, respectively. When presenting the algorithm, we treat $j$ as the site index, with $j=1,\ldots,N$. For cases where fermions have internal degrees of freedom (e.g., spin or orbital), one may view $j$ as the combination of site and internal state indices, e.g., $j=(i,\sigma)$.

Let us consider a pure U(1)-FGS $|\psi\rangle$, which is fully characterized by its correlation matrix~\cite{Peschel2003,Bravyi2005}
\begin{align}
\mathcal{C}_{j,j'}= 2 \,  \langle \psi| c^{\dagger}_{j} c_{j'} |\psi\rangle -\delta_{j,j'} \, .
\label{eq:CM}
\end{align}
$\mathcal{C}$ is an $N \times N$ Hermitian matrix satisfying $\mathcal{C}^{2} = \mathbbm{1}_{N}$. Diagonalizing the correlation matrix $\mathcal{C}$ gives 
\begin{align}
    M^{\dag}\mathcal{C} M = \left( \begin{matrix} \mathbbm{1}_{Q} &  0 \\ 0 & -\mathbbm{1}_{N-Q} \end{matrix} \right) ,
\label{eq:CM-diagonalization}
\end{align}
where $M$ is a unitary matrix. $Q$ is the number of occupied modes in $|\psi\rangle$, and the corresponding eigenvectors (first $Q$ rows of the unitary matrix $M^{\dag}$) can be used to write $|\psi\rangle$ explicitly as a Fermi sea state
\begin{align}
|\psi\rangle = \prod_{q=1}^{Q} f^{\dagger}_{q} |0\rangle \, ,
\label{eq:Fermi-sea}
\end{align}
where $f^{\dag}_q = \sum_{j=1}^{N} M^{\dag}_{q,j}c^{\dag}_{j}$ with $q=1,\ldots,Q$ ($q=Q+1,\ldots,N$) are occupied (unoccupied) modes in $|\psi\rangle$, and $|0\rangle$ is the vacuum, $c_{j}|0\rangle=0 \; \forall j$.

For our algorithm, we need the Schmidt decomposition of U(1)-FGSs. Let us partition the first $N_{\mathrm{L}}$ sites (index $j=1,\ldots,N_{\mathrm{L}}$) of the system into the left part (denoted as ``L'') and remaining $N_{\mathrm{R}}=N-N_{\mathrm{L}}$ sites into the right part (denoted as ``R''). Accordingly, the correlation matrix in Eq.~\eqref{eq:CM} is decomposed into four blocks
\begin{align}
    \mathcal{C} =  \begin{pmatrix} \mathcal{C}^{\mathrm{LL}} & \mathcal{C}^{\mathrm{LR}}  \\ (\mathcal{C}^{\mathrm{LR}})^{\dag} & \mathcal{C}^{\mathrm{RR}} \end{pmatrix}  ,
\label{eq:CM-decomposition}
\end{align}
where $\mathcal{C}^{\mathrm{LL}}$ and $\mathcal{C}^{\mathrm{RR}}$ are $N_{\mathrm{L}} \times N_{\mathrm{L}}$ and $N_{\mathrm{R}} \times N_{\mathrm{R}}$ Hermitian matrices, respectively. Actually, $\mathcal{C}^{\mathrm{LL}}$ is the correlation matrix of the reduced density operator $\rho^{\mathrm{L}}$ for the left subsystem
\begin{align}
\mathcal{C}^{\mathrm{LL}}_{j,j'} = 2 \,  \mathrm{tr}_{\mathrm{L}}(\rho^{\mathrm{L}} c^{\dagger}_{j} c_{j'})-\delta_{j,j'}
\end{align}
with $1\leq j,j^{\prime}\leq N_{\mathrm{L}}$ and ``$\mathrm{tr}_{\mathrm{L}}$'' being the trace over degrees of freedom in the left subsystem. Similar definition holds for $\mathcal{C^{\mathrm{RR}}}$ in the right subsystem.

Substituting Eq.~\eqref{eq:CM-decomposition} into $\mathcal{C}^{2} = \mathbbm{1}_{N}$ gives $(\mathcal{C}^{\mathrm{LL}})^2 + \mathcal{C}^{\mathrm{LR}}(\mathcal{C}^{\mathrm{LR}})^{\dag} = \mathbbm{1}_{N_{\mathrm{L}}}$ and $(\mathcal{C}^{\mathrm{LR}})^{\dag} \mathcal{C}^{\mathrm{LR}} + (\mathcal{C}^{\mathrm{RR}})^2 = \mathbbm{1}_{N-N_{\mathrm{L}}}$. It becomes clear that the two unitary matrices obtained from the singular value decomposition (SVD) of $\mathcal{C}^{\mathrm{LR}}$ can diagonalize $\mathcal{C}^{\mathrm{LL}}$ and $\mathcal{C^{\mathrm{RR}}}$. For instance, we have
\begin{align}
U^{\dag} \mathcal{C}^{\mathrm{LL}} U
&= \begin{pmatrix}
\Lambda &  0 &  0 \\
0 & \mathbbm{1} & 0 \\
0 & 0 & -\mathbbm{1}
\end{pmatrix} \, ,
\label{eq:sub-CM-diagonalization-1}
\end{align}
where $\Lambda$ is a diagonal matrix with diagonal entries $\Lambda_{\mu} \in (-1,1)$. The size of the three diagonal blocks in Eq.~\eqref{eq:sub-CM-diagonalization-1} depends on the entanglement structure of $|\psi\rangle$. We shall denote these blocks as ``occupied'', ``entangled'', and ``unoccupied'' (abbreviated as $\mathrm{O}$, $\mathrm{E}$, and $\mathrm{U}$, respectively), for the reason that will become clear soon. The unitary matrix $U$ defines a single-particle basis rotation in the left subsystem
\begin{align}
d^{\dag}_{\mathrm{L},\mu} = \sum_{j=1}^{N_{\mathrm{L}}} U^{\dag}_{\mu,j} c^{\dag}_{j}
\label{eq:subsystem-orbital-1}
\end{align}
with $\mu = 1,\ldots,N_{\mathrm{L}}$. In this new basis, the three blocks in the diagonalized subsystem correlation matrix [Eq.~\eqref{eq:sub-CM-diagonalization-1}] translates into
\begin{align}
\langle \psi| d^{\dag}_{\mathrm{L},\mu} d_{\mathrm{L},\mu} |\psi\rangle = 
\begin{cases}
    1 & \text{if $\mu \in \mathrm{O}$} \\
    (1+\Lambda_{\mu})/2 & \text{if $\mu \in \mathrm{E}$} \\
    0 & \text{if $\mu \in \mathrm{U}$}
\end{cases},
\label{eq:L-occupation}
\end{align}
where $\mu \in \mathrm{O},\mathrm{E},\mathrm{U}$ distinguish three type of modes: $d^{\dag}_{\mathrm{L},\mu \in \mathrm{O}}$ ($d^{\dag}_{\mathrm{L},\mu \in \mathrm{U}}$) modes are completely occupied (unoccupied) in $|\psi\rangle$ and have no entanglement with the right subsystem, and $d^{\dag}_{\mathrm{L},\mu \in \mathrm{E}}$ are only partially occupied in $|\psi\rangle$ and hence entangle with certain modes in the right subsystem. In this context, $\mathrm{O},\mathrm{E},\mathrm{U}$ should not be viewed as particular sets; they are rather labels for modes (we will use them for the right subsystem, too).

Similarly, the diagonalization of $\mathcal{C}^{\mathrm{RR}}$ gives
\begin{align}
V^{\dag} \mathcal{C}^{\mathrm{RR}} V
&= \begin{pmatrix}
-\Lambda &  0 &  0 \\
0 & \mathbbm{1} & 0 \\
0 & 0 & -\mathbbm{1}
\end{pmatrix} \, ,
\label{eq:sub-CM-diagonalization-2}
\end{align}
where the unitary matrix $V$ defines an orthonormal single-particle basis for the right subsystem
\begin{align}
d^{\dag}_{\mathrm{R},\nu} = \sum_{j=N_{\mathrm{L}}+1}^{N} V^{\dag}_{\nu,j} c^{\dag}_{j}
\label{eq:subsystem-orbital-2}
\end{align}
with $\nu = 1,\ldots,N_{\mathrm{R}}$. The occupation of these modes in $|\psi\rangle$ is given by
\begin{align}
\langle \psi| d^{\dag}_{\mathrm{R},\nu} d_{\mathrm{R},\nu} |\psi\rangle = 
\begin{cases}
    1 & \text{if $\nu \in \mathrm{O}$} \\
    (1-\Lambda_{\nu})/2 & \text{if $\nu \in \mathrm{E}$} \\
    0 & \text{if $\nu \in \mathrm{U}$}
\end{cases}.
\label{eq:R-occupation}
\end{align}
Combining that the unitary matrices $U^{\dag}$ and $V$ SVD-diagonalize $\mathcal{C}^{\mathrm{LR}}$ [i.e., $(U^{\dag} \mathcal{C}^{\mathrm{LR}}V)_{\mu,\nu} = \langle \psi| d^{\dag}_{\mathrm{L},\mu} d_{\mathrm{R},\nu} |\psi\rangle \propto \delta_{\mu,\nu} $] and that $\mathcal{C}^{2} = \mathbbm{1}_{N}$ (valid also in the new basis), we obtain
\begin{align}
    \langle \psi| d^{\dag}_{\mathrm{L},\mu} d_{\mathrm{R},\nu} |\psi\rangle = \frac{\sqrt{1-\Lambda^2_{\mu}}}{2} \delta_{\mu,\nu}
\end{align}
for $\mu,\nu \in \mathrm{E}$. Together with Eqs.~\eqref{eq:L-occupation} and \eqref{eq:R-occupation}, we can reconstruct $|\psi\rangle$ in terms of the orthonormal modes defined in two subsystems as follows:
\begin{align}
    |\psi\rangle &= \prod_{\mu \in \mathrm{E}} \left[ \sqrt{\frac{1+\Lambda_{\mu}}{2}} d^{\dagger}_{\mathrm{L},\mu} + \sqrt{\frac{1-\Lambda_{\mu}}{2}} d^{\dagger}_{\mathrm{R},\mu} \right] \nonumber \\
    &\phantom{=} \quad \times \prod_{\mu \in \mathrm{O}} d^{\dagger}_{\mathrm{L},\mu} \prod_{\nu \in \mathrm{O}} d^{\dagger}_{\mathrm{R},\nu} |0\rangle \, .
\label{eq:mode-decomposition}
\end{align}
When multiplying out the product for entangled modes ($\mu \in \mathrm{E}$), this gives the Schmidt decomposition of $|\psi\rangle$.

\begin{figure}
    \centering
    \includegraphics[width=1.0\linewidth]{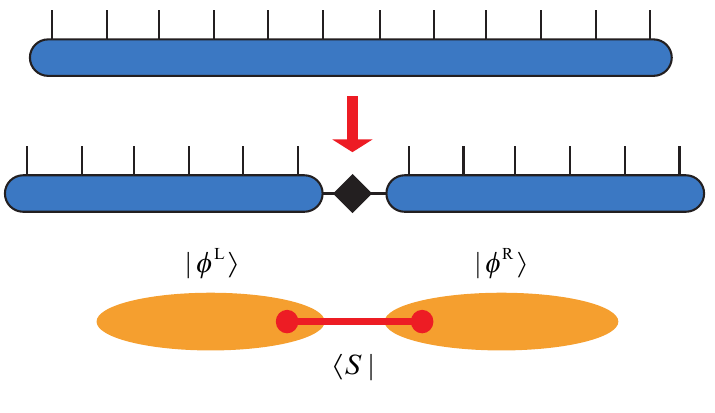}
    \caption{Schematic of the Schmidt decomposition of a fermionic Gaussian state $|\psi\rangle$ via introducing virtual fermionic modes and the associated virtual bond state $\langle S|$.}
    \label{fig:Schmidt-decomposition}
\end{figure}

For later discussions, it will be useful to encode the Schmidt vectors in two U(1)-FGSs living in enlarged Hilbert spaces
\begin{align}
|\phi^{\mathrm{L}}\rangle &= \prod_{\mu=1}^{N_{\mathrm{L}}} \left[ \frac{1}{\sqrt{2}}\left(d^{\dagger}_{\mathrm{L},\mu}+b^{\dagger}_{N_{\mathrm{L}},\mu}\right)\right] |0\rangle_{\mathrm{p,v}},  \nonumber \\
|\phi^{\mathrm{R}}\rangle &= \prod_{\nu=1}^{N_{\mathrm{R}}} \left[\frac{1}{\sqrt{2}}\left(d^{\dagger}_{\mathrm{R},\nu}+b^{\dagger}_{N_{\mathrm{L}+1},\nu}\right)\right] |0\rangle_{\mathrm{p,v}},
\label{eq:encoded-Schmidt-vectors}
\end{align}
where every subsystem orthonormal mode ($d^{\dagger}_{\mathrm{L},\mu}$ and $d^{\dagger}_{\mathrm{R},\nu}$) entangles with one virtual mode ($b^{\dagger}_{N_\mathrm{L},\mu}$  and $b^{\dagger}_{N_{\mathrm{L}}+1,\nu}$). $|0\rangle_{\mathrm{p,v}}$ is the shared vacuum of physical (``p'') and virtual (``v'') modes. The virtual modes, indicated by their site indices $N_\mathrm{L}$ and $N_{\mathrm{L}+1}$, locate at the ``entangling edge'' of the Schmidt decomposition. Introducing a virtual bond state between virtual modes
\begin{align}
|S\rangle &= \prod_{\mu \in \mathrm{E}} \left[ \sqrt{\frac{1-\Lambda_{\mu}}{2}} b^{\dagger}_{N_{\mathrm{L}},\mu} - \sqrt{\frac{1+\Lambda_{\mu}}{2}} b^{\dagger}_{N_{\mathrm{L}}+1,\mu} \right]   \nonumber \\
&\phantom{=} \quad \times \prod_{\mu \in \mathrm{U}} b^{\dagger}_{N_{\mathrm{L}},\mu} \prod_{\nu \in \mathrm{U}} b^{\dagger}_{N_{\mathrm{L}}+1,\nu} |0\rangle_{\mathrm{v}} \, ,
\label{eq:S}
\end{align}
the original state $|\psi\rangle$ is restored by contracting the virtual modes (depicted in Fig.~\ref{fig:Schmidt-decomposition}):
\begin{align}
|\psi\rangle = \langle S| \left(|\phi^{\mathrm{L}}\rangle \otimes |\phi^{\mathrm{R}}\rangle\right) \, ,
\label{eq:Schmidt-bond-form}
\end{align}
where an unimportant overall factor is ignored. This form of the Schmidt decomposition will be used in Sec.~\ref{subsec:methodB}.
As a minor remark, the tensor product notation in Eq.~\eqref{eq:Schmidt-bond-form} is not entirely rigorous in mathematical sense, since fermionic systems need a $\mathbb{Z}_2$-graded tensor product for suitably bookkeeping the fermionic signs. Nevertheless, for our purpose, this notation is intuitive and does not cause sign inconsistency. Below we shall continue with such notations (and sometimes even drop the tensor product symbol).

\subsection{Calculation of the MPS local correlation matrix}
\label{subsec:methodB}

In this subsection, we illustrate how to obtain an MPS representation of $|\psi\rangle$. The elementary building blocks of MPSs are single-site tensors. The key part of our algorithm is the calculation of a central-site tensor of the MPS. For finite-size systems without translation invariance, MPS tensors at other lattice sites can be obtained by iterating this algorithm. For one-site translation invariant systems, one can gradually increase system sizes such that the central-site tensor is converged; this gives an iMPS (by copying the single-site tensor to every site) in the thermodynamic limit. For systems with multi-site unit cells (e.g., cylinders), iMPSs can be obtained by calculating single-site tensors separately for all sites within one unit cell.

Let us choose a central site with index $m$ ($m \sim N/2$). We first follow the procedure in Sec.~\ref{subsec:methodA} and decompose $|\psi\rangle$ in the form of Eq.~\eqref{eq:Schmidt-bond-form}:
\begin{align}
|\psi\rangle = \langle S_{m,m+1}| \left(|\phi^{[\leq m]}\rangle \otimes |\phi^{[>m]}\rangle \right),
\label{eq:recursion-1}
\end{align}
where $\langle S_{m,m+1}|$ is the virtual bond state, and $|\phi^{[\leq m]}\rangle$ and $|\phi^{[>m]}\rangle$) are U(1)-FGSs encoding the Schmidt vectors of the left (sites $1$ to $m$) and right (sites $m+1$ to $N$) subsystems, respectively. For the derivations below, we only need $|\phi^{[\leq m]}\rangle$:
\begin{align}
|\phi^{[\leq m]}\rangle &= \prod_{\mu=1}^{m} \left[\frac{1}{\sqrt{2}} \left(d^{\dagger}_{[\leq m],\mu} + b^{\dagger}_{m,r,\mu} \right)\right] |0\rangle_{\mathrm{p,v}} \, ,
\label{eq:Schmidt-vector-1}
\end{align}
where $d^{\dagger}_{[\leq m],\mu}$ are orthonormal modes for the left subsystem [see Eq.~\eqref{eq:subsystem-orbital-1}], and  $b^{\dagger}_{m,r,\mu}$ are virtual modes on site $m$ (index ``$r$'' means right side), maximally entangling with $d^{\dagger}_{[\leq m],\mu}$ [see Eq.~\eqref{eq:encoded-Schmidt-vectors}].

\begin{figure}
    \centering
    \includegraphics[width=1.0\linewidth]{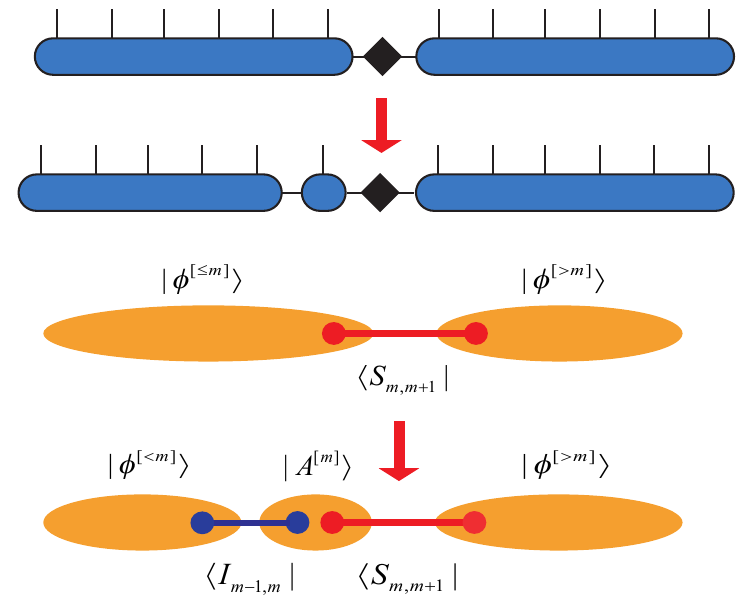}
    \caption{Further decomposition of the fermionic Gaussian state $|\psi\rangle$ to obtain the local state $|A^{[m]}\rangle$ at site $m$.}
    \label{fig:further-decomposition}
\end{figure}

\begin{figure}
    \centering
    \includegraphics[width=1.0\linewidth]{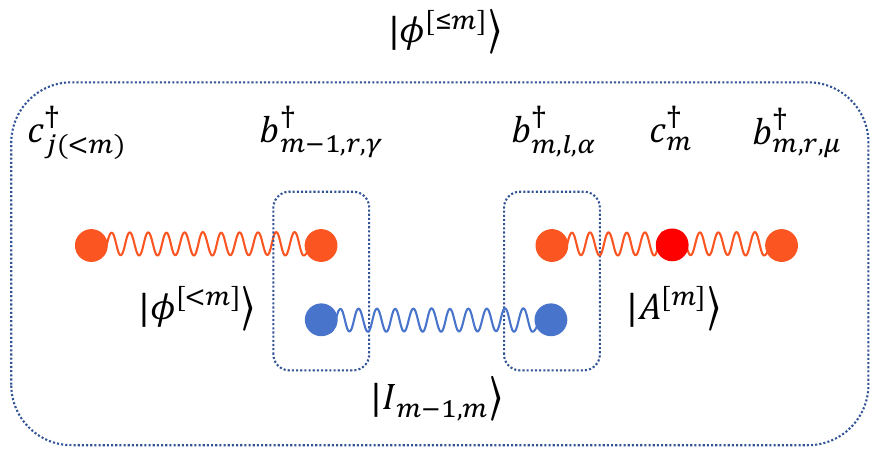}
    \caption{Physical and virtual modes forming the vectors $|\phi^{[\leq m]}\rangle$, $|\phi^{[<m]}\rangle$, $|A^{[m]}\rangle$, and $|I_{m-1,m}\rangle$ in Eq.~\eqref{eq:recursion-2}.}
\label{fig:decomposed-vectors}
\end{figure}

To get a local state at site $m$ (playing the role of an MPS tensor), we decompose $|\psi\rangle$ further as
\begin{align}
|\psi\rangle &= \left(\langle I_{m-1,m}|\otimes \langle S_{m,m+1}|\right) \nonumber \\
&\phantom{=} \quad \times \left(|\phi^{[<m]}\rangle \otimes |A^{[m]}\rangle \otimes |\phi^{[>m]}\rangle \right)
\label{eq:further-decomposition-1}
\end{align}
by requiring
\begin{align}
|\phi^{[\leq m]}\rangle = \langle I_{m-1,m}| \left( |\phi^{[<m]}\rangle \otimes |A^{[m]}\rangle \right)
\label{eq:recursion-2}
\end{align}
in Eq.~\eqref{eq:recursion-1}. Here $|\phi^{[<m]}\rangle$ is the state encoding the Schmidt vectors of the new left subsystem (including sites $1$ to $m-1$)
\begin{align}
|\phi^{[<m]}\rangle = \prod_{\gamma=1}^{m-1} \left[ \frac{1}{\sqrt{2}}\left(d^{\dagger}_{[<m],\gamma}+b^{\dagger}_{m-1,r,\gamma}\right)\right]|0\rangle_{\mathrm{v}} \, ,
\label{eq:Schmidt-vector-2}
\end{align}
and $|I_{m-1,m}\rangle$ is a maximally entangled virtual bond state between sites $m-1$ and $m$:
\begin{align}
|I_{m-1,m}\rangle = \prod_{\alpha=1}^{m-1} \left[ \frac{1}{\sqrt{2}}\left(b^{\dagger}_{m-1,r,\alpha} + b^{\dagger}_{m,l,\alpha}\right) \right]|0\rangle_{\mathrm{v}} \, ,
\label{eq:entangled-bond}
\end{align}
where $b^{\dagger}_{m,l,\alpha}$ are virtual modes on site $m$ (index ``$l$'' means left side). The state $|A^{[m]}\rangle$, appearing in Eq.~\eqref{eq:recursion-2}, is a U(1)-FGS at site $m$, which consists of the physical mode $c^{\dag}_m$ and virtual modes $b^{\dagger}_{m,l,\alpha}$ [$\alpha=1,\ldots,(m-1)$] and $b^{\dagger}_{m,r,\mu}$ ($\mu=1,\ldots,m$). The explicit form of $|A^{[m]}\rangle$ needs to be determined. The graphical representation of the decompositions in Eqs.~\eqref{eq:recursion-1}, \eqref{eq:further-decomposition-1} and \eqref{eq:recursion-2} is depicted in Fig.~\ref{fig:further-decomposition}. 
To determine $|A^{[m]}\rangle$, we observe that a result in Ref.~\cite{LiJW2023} [Eq.~(A14) thereof] can be used to translate Eq.~\eqref{eq:recursion-2} into a relation of correlation matrices of the U(1)-FGSs involved, from which the correlation matrix of $|A^{[m]}\rangle$ can be calculated. Let us consider an input pure U(1)-FGS $|\psi^{in}\rangle$ and divide the whole system into two subsystems denoted as $A$ and $B$. Its correlation matrix can be decomposed to the block form
\begin{align}
\mathcal{C}^{\mathrm{in}}=\begin{pmatrix}
    \mathcal{C}^{AA} & \mathcal{C}^{AB} \\
    (\mathcal{C}^{AB})^{\dagger} & \mathcal{C}^{BB}
\end{pmatrix},
\end{align}
where the superscripts indicate the degrees of freedom residing in the two subsystems. After projecting the subsystem $B$ onto a pure U(1)-FGS with correlation matrix $\mathcal{C}^B$, we obtain a U(1)-FGS $|\psi^{\mathrm{out}}\rangle$ with correlation matrix
\begin{align}
\mathcal{C}^{\mathrm{out}}=\mathcal{C}^{AA}-\mathcal{C}^{AB}(\mathcal{C}^{BB}+\mathcal{C}^I)^{-1}(\mathcal{C}^{AB})^{\dagger}
\end{align}
Before using this result, one should first decompose the correlation matrices into blocks corresponding to certain modes (see Fig.~\ref{fig:decomposed-vectors}). For instance, the correlation matrix of $|A^{[m]}\rangle$, denoted by $\mathcal{C}^{[m]}$, takes a $3 \times 3$ block form
\begin{align}
\mathcal{C}^{[m]}=
    \begin{pNiceMatrix}[first-row,first-col]
	&  \scriptstyle{(m,l,\alpha^{\prime})}  & \scriptstyle {(m)}  & \scriptstyle{(m,r,\mu^{\prime})} \\[2mm]
	  \scriptstyle {(m, l, \alpha)} & \mathcal{C}^{[m]}_{1,1} & \mathcal{C}^{[m]}_{1,2} & \mathcal{C}^{[m]}_{1,3}\\[2mm]
	  \scriptstyle {(m)} & [\mathcal{C}^{[m]}_{1,2}]^{\dagger} & \mathcal{C}^{[m]}_{2,2} & \mathcal{C}^{[m]}_{2,3}\\[2mm]
   	\scriptstyle {(m, r, \mu)} & [\mathcal{C}^{[m]}_{1,3}]^{\dagger} & [\mathcal{C}^{[m]}_{2,3}]^{\dagger} & \mathcal{C}^{[m]}_{3,3} 
    \end{pNiceMatrix}
\label{eq:A-CM}
\end{align}
with 
\begin{align}
(\mathcal{C}^{[m]}_{1,1})_{(m,l,\alpha),(m,l,\alpha^{\prime})} &= 2\langle A^{[m]}| b^{\dagger}_{m,l,\alpha } b_{m,l,\alpha^{\prime}}|A^{[m]}\rangle -\delta_{\alpha,\alpha^{\prime}},  \nonumber \\
(\mathcal{C}^{[m]}_{2,2})_{(m),(m)} &= 2\langle A^{[m]}| c^{\dagger}_{m} c_{m}|A^{[m]}\rangle - 1,  \nonumber \\
(\mathcal{C}^{[m]}_{3,3})_{(m,r,\mu),(m,r,\mu^{\prime})} &= 2\langle A^{[m]}| b^{\dagger}_{m,r,\mu } b_{m,r,\mu^{\prime}}|A^{[m]}\rangle -\delta_{\mu,\mu^{\prime}},   \nonumber \\
(\mathcal{C}^{[m]}_{1,2})_{(m,l,\alpha),(m)} &= 2\langle A^{[m]}| b^{\dagger}_{m,l,\alpha} c_{m}|A^{[m]}\rangle,  \nonumber \\ 
(\mathcal{C}^{[m]}_{1,3})_{(m,l,\alpha),(m,r,\mu^{\prime})} &= 2\langle A^{[m]}| b^{\dagger}_{m,l,\alpha } b_{m,r,\mu^{\prime}}|A^{[m]}\rangle,   \nonumber \\
(\mathcal{C}^{[m]}_{2,3})_{(m),(m,r,\mu^{\prime})} &= 2\langle A^{[m]}| c^{\dagger}_{m} b_{m,r,\mu^{\prime}}|A^{[m]}\rangle.
\end{align}
For ease of reading, we have indicated the blocks in Eq.~\eqref{eq:A-CM} with the corresponding indices of the fermionic modes (this will also be used for other correlation matrices below). For rows and columns in Eq.~\eqref{eq:A-CM} involving the physical mode, $(m)$ is an index with dimension one (physical modes are ``spinless''); nevertheless, we keep it to have a uniform format (it will be convenient for extensions to cases where physical modes do have internal degrees of freedom).

The correlation matrix of $|\phi^{[<m]}\rangle$ [Eq.~\eqref{eq:Schmidt-vector-2}], denoted by $\mathcal{C}^{[<m]}$, has a $2 \times 2$ block form
\begin{align}
\mathcal{C}^{[<m]}=
    \begin{pNiceMatrix}[first-row,first-col]
	  &  \scriptstyle {(j^{\prime}\textless m)} & \scriptstyle{(m-1,r,\gamma^{\prime})} \\[2mm]
	  \scriptstyle {(j\textless m)} & 0 & \mathcal{C}^{[<m]}_{1,2} \\[2mm]
	  \scriptstyle {(m-1, r, \gamma)} & [\mathcal{C}^{[<m]}_{1,2}]^{\dagger} & 0 
    \end{pNiceMatrix},
\label{eq:C1-CM}
\end{align}
where $\mathcal{C}^{[<m]}_{1,2}$ has entries $(\mathcal{C}^{[<m]}_{1,2})_{(j),(m-1,r,\gamma^{\prime})} = 2 \langle \phi^{[<m]}| c^{\dagger}_{j} b_{m-1,r,\gamma^{\prime}}|\phi^{[<m]}\rangle$. Here $j$ and $\gamma^{\prime}$ range from $1$ to $(m-1)$. Following Eq.~\eqref{eq:subsystem-orbital-1}, we write the subsystem orthonormal modes as  $d^{\dag}_{[<m],\gamma} = \sum_{j=1}^{m-1} (U^{[<m]})^{\dag}_{\gamma,j}c^{\dag}_{j}$, where the unitary matrix $U^{[<m]}$ diagonalizes the subsystem correlation matrix from site $1$ to $(m-1)$ [c.f. Eq.~\eqref{eq:sub-CM-diagonalization-1}]. The definition of $|\phi^{[<m]}\rangle$ in Eq.~\eqref{eq:Schmidt-vector-2} gives $2\langle \phi^{[<m]}| d^{\dagger}_{[<m],\gamma} b_{m-1,r,\gamma^{\prime}}|\phi^{[<m]}\rangle
= \delta_{\gamma,\gamma^{\prime}}$, which leads to
\begin{align}
\mathcal{C}^{[<m]}_{1,2} = U^{[<m]} .
\label{eq:C1-CM-sub}
\end{align}
Moreover, we obtain $2\langle \phi^{[<m]}| c^{\dagger}_{j} c_{j^{\prime}}|\phi^{[<m]}\rangle -\delta_{j,j^{\prime}} = 2\langle \phi^{[<m]}| b^{\dagger}_{m-1,r,\gamma} b_{m-1,r,\gamma^{\prime}}|\phi^{[<m]}\rangle -\delta_{\gamma,\gamma^{\prime}} = 0$, which correspond to the vanishing diagonal blocks in Eq.~\eqref{eq:C1-CM}. 

The correlation matrix of $|\phi^{[\leq m]}\rangle$ [Eq.~\eqref{eq:Schmidt-vector-1}] is given by
\begin{align}
\mathcal{C}^{[\leq m]}=
    \begin{pNiceMatrix}[first-row,first-col]
	  &  \scriptstyle {(j^{\prime}\textless m)} & \scriptstyle {(m)} & \scriptstyle{(m,r,\mu^{\prime})} \\[2mm] 
	  \scriptstyle {(j\textless m)} & 0 & 0 & \mathcal{C}^{[\leq m]}_{1,3} \\[2mm]
        \scriptstyle {(m)} & 0 & 0 & \mathcal{C}^{[\leq m]}_{2,3} \\[2mm]
        \scriptstyle {( m, r, \mu)} & [\mathcal{C}^{[\leq m]}_{1,3}]^{\dagger} & [\mathcal{C}^{[\leq m]}_{2,3}]^{\dagger} &0
    \end{pNiceMatrix}
\label{eq:C2-CM}
\end{align}
with $(\mathcal{C}^{[\leq m]}_{1,3})_{(j),(m,r,\mu^{\prime})} = 2\langle \phi^{[\leq m]}| c^{\dagger}_{j} b_{m,r,\mu^{\prime}}|\phi^{[\leq m]}\rangle$ and $(\mathcal{C}^{[\leq m]}_{2,3})_{(m),(m,r,\mu^{\prime})} = 2\langle \phi^{[\leq m]}| c^{\dagger}_{m} b_{m,r,\mu^{\prime}}|\phi^{[\leq m]}\rangle$, where $j$ ranges from $1$ to $(m-1)$ and $\mu^{\prime}$ from $1$ to $m$. The remaining derivation is very similar to that for $|\phi^{[<m]}\rangle$ above and will not be repeated. If we write the subsystem orthonormal modes as $d^{\dag}_{[\leq m],\mu} = \sum_{j=1}^{m} (U^{[\leq m]})^{\dag}_{\mu,j}c^{\dag}_{j} \equiv \sum_{j=1}^{m-1} (U_1^{[\leq m]})^{\dag}_{\mu,j}c^{\dag}_{j} + (U_2^{[\leq m]})^{\dag} c^{\dag}_{m}$, we obtain
\begin{align}
\begin{pmatrix}
    \mathcal{C}^{[\leq m]}_{1,3} \\
    \mathcal{C}^{[\leq m]}_{2,3}
\end{pmatrix}
 = U^{[\leq m]} = 
 \begin{pmatrix}
    U_1^{[\leq m]} \\
    U_2^{[\leq m]}
\end{pmatrix},
\label{eq:C2-CM-sub}
\end{align}
where $U_1^{[\leq m]}$ ($U_2^{[\leq m]}$) is an $(m-1)\times m$ matrix (a $1 \times m$ row vector).

The correlation matrix of the virtual bond state $|I_{m-1,m}\rangle$ [Eq.~\eqref{eq:entangled-bond}] is most easily obtained
\begin{align}
\mathcal{C}^{I}=
    \begin{pNiceMatrix}[first-row,first-col]
	  &  \scriptstyle {(m-1,r,\gamma^{\prime})} & \scriptstyle{(m,l,\alpha^{\prime})} \\[2mm]
	  \scriptstyle {(m-1,r,\gamma)} & 0 & \mathbbm{1}_{m-1} \\[2mm]
	  \scriptstyle {(m, l, \alpha)} & \mathbbm{1}_{m-1} & 0 
    \end{pNiceMatrix},
\label{eq:CI-CM}
\end{align}
where the non-vanishing off-diagonal block comes from $2\langle I_{m-1,m}| b^{\dagger}_{m-1,r,\gamma} b_{m,l,\alpha^{\prime}}|I_{m-1,m}\rangle
= \delta_{\gamma,\alpha^{\prime}}$.

By using Eq.~(A14) in Ref.~\cite{LiJW2023}, Eq.~\eqref{eq:recursion-2} translates into the following correlation matrix equation:
\begin{align}
\label{eq:recursion-3}
&\begin{pmatrix}  
0 & 0 & \mathcal{C}^{[\leq m]}_{1,3} \\
0 & 0 & \mathcal{C}^{[\leq m]}_{2,3} \\
[\mathcal{C}^{[\leq m]}_{1,3}]^{\dagger} & [\mathcal{C}^{[\leq m]}_{2,3}]^{\dagger} &0
\end{pmatrix}=
\begin{pmatrix}
0 & 0 & 0 \\
0 & \mathcal{C}^{[m]}_{2,2} & \mathcal{C}^{[m]}_{2,3} \\
0 & [\mathcal{C}^{[m]}_{2,3}]^{\dagger} & \mathcal{C}^{[m]}_{3,3}
\end{pmatrix} \nonumber \\
-&\begin{pmatrix}
\mathcal{C}^{[<m]}_{1,2} & 0  \\
0 & [\mathcal{C}^{[m]}_{1,2}]^{\dagger}  \\
0 & [\mathcal{C}^{[m]}_{1,3}]^{\dagger} 
\end{pmatrix}
\begin{pmatrix}
0 &  \mathbbm{1} \\
\mathbbm{1} & \mathcal{C}^{[m]}_{1,1} 
\end{pmatrix}^{-1}
\begin{pmatrix}
[\mathcal{C}^{[<m]}_{1,2}]^{\dagger} &  0 & 0 \\
0  & \mathcal{C}^{[m]}_{1,2} & \mathcal{C}^{[m]}_{1,3}
\end{pmatrix},
\end{align}
from which the solution for the correlation matrix $\mathcal{C}^{[m]}$ of the local state $|A^{[m]}\rangle$ reads
\begin{align}
&\mathcal{C}^{[m]}_{1,1}=0, \quad \mathcal{C}^{[m]}_{1,2}=0, \quad \mathcal{C}^{[m]}_{2,2}=0, \quad \mathcal{C}^{[m]}_{3,3}=0,  \nonumber \\
&\mathcal{C}^{[m]}_{1,3}=-[\mathcal{C}^{[<m]}_{1,2}]^{\dagger}C^{[\leq m]}_{1,3}, \quad
\mathcal{C}^{[m]}_{2,3}=C^{[\leq m]}_{2,3}.
\end{align}
Together with Eqs.~\eqref{eq:C1-CM-sub} and \eqref{eq:C2-CM-sub}, the solution for Eq.~\eqref{eq:A-CM} is given by
\begin{align}
\mathcal{C}^{[m]}=
\begin{pmatrix}
0 & 0 & -(U^{[<m]})^{\dagger} U_{1}^{[\leq m]} \\
0 & 0 & U_{2}^{[\leq m]} \\
-(U_{1}^{[\leq m]})^{\dagger} U^{[<m]} & (U_{2}^{[\leq m]})^{\dagger} & 0 
\end{pmatrix}.
\label{eq:A-CM-solution}
\end{align}
Thus, once the unitary matrices diagonalizing subsystem correlation matrices are determined, $\mathcal{C}^{[m]}$ as the local correlation matrix of the MPS is also obtained.

\subsection{Calculation of the MPS local tensor}
\label{subsec:methodC}

With the local correlation matrix $\mathcal{C}^{[m]}$ in Eq.~\eqref{eq:A-CM-solution}, we now demonstrate how to obtain a concrete form of the its corresponding Gaussian state $|A^{[m]}\rangle$ and write it explicitly as an MPS local tensor. As the correlation matrix of a U(1)-FGS, $\mathcal{C}^{[m]}$ can be diagonalized as
\begin{align}
S^{\dagger}\mathcal{C}^{[m]} S = 
\begin{pmatrix}
\mathbbm{1}_{Q_m} &  0 \\
0 & -\mathbbm{1}_{2m-Q_m}
\end{pmatrix},
\label{eq:diagonal}
\end{align}
where the total number of local modes (including both virtual and physical modes) in $|A^{[m]}\rangle$ is $2m$ and $S$ is a unitary matrix. Then, we can write $|A^{[m]}\rangle$ as
\begin{align}
|A^{[m]}\rangle=\prod_{p=1}^{Q_m}d^{\dagger}_p |0\rangle_{\mathrm{p,v}}, 
\label{eq:local-Am-1}
\end{align}
where the $Q_m$ occupied modes are given by
\begin{align}
d^{\dagger}_{p}=S^{\dagger}_{p,0} c^{\dagger}_{m}+ \sum_{\alpha=1}^{m-1} S^{\dagger}_{p,(l,\alpha)} b^{\dagger}_{m,l,\alpha}
+\sum_{\beta=1}^{m} S^{\dagger}_{p,(r,\beta)} b^{\dagger}_{m,r,\beta} .
\label{eq:occupied-d-mode}
\end{align}
The ``column index'' $0$ in $S^{\dagger}_{p,0}$ is used for denoting the local physical mode. If we rewrite Eq.~\eqref{eq:local-Am-1} in the Fock basis of physical and virtual modes, it becomes
\begin{align}
|A^{[m]}\rangle &= \sum_{ \{n_0,n_{l,\alpha},n_{r,\beta}\} }  A^{\{n_{0}\}}_{\{n_{l,\alpha}\},\{n_{r,\beta}\}}   \nonumber \\
&\phantom{=} \quad \times (c^{\dagger}_{m})^{n_0}(b^{\dagger}_{m,l,\alpha})^{n_{l,\alpha}}(b^{\dagger}_{m,r,\beta})^{n_{r,\beta}}|0\rangle_{\mathrm{p,v}},
\label{eq:local-Am-2}
\end{align}
where $n_0$, $n_{l,\alpha}$ and $n_{r,\beta}$ are occupation numbers of the local physical, left virtual, and right virtual modes, respectively. Here, we have used a binary list $\{n_0\},{\{n_{l,\alpha}\},\{n_{r,\beta}\}}$ to label the fermionic states. For instance, \{0101\} represents a state with only the second and the fourth modes occupied. The coefficient $A^{\{n_{0}\}}_{\{n_{l,\alpha}\},\{n_{r,\beta}\}}$ in Eq.~\eqref{eq:local-Am-2} is the Slater determinant defined with $S[\{n_0\},\{n_{l,\alpha}\},\{n_{r,\beta}\}]$, which is the sub-matrix taken form the unitary matrix $S$ by choosing its first $Q_m$ rows and the columns chosen with respect to occupied physical and virtual fermionic modes. The coefficients for virtual states, such as $|I_{m,m+1}\rangle$ and $|S_{m,m+1}\rangle$, can be calculated in the same way.

The main bottleneck for this step is that for large systems, converting the local correlation matrix into a local tensor (or calculating all the coefficients) is not manageable, as the bond dimension would grow exponentially. For obtaining an MPS with a feasible bond dimension $D$ ($D = 10^3 \sim 10^4 $), we keep only $D$ states of virtual fermions, corresponding to $D$ elements in binary lists $\{n_{l,\alpha}\}$ and $\{n_{r,\beta}\}$. The selection criterion for the kept elements is that they have dominant contributions to the bipartite entanglement according to the Schmidt decompsition in Eq.~\eqref{eq:S} (this is equivalent to the SVD truncation~\cite{Petrica2021}). If SVD truncation is performed on a Gaussian state, the result is no longer Gaussian so the Schmidt weights cannot be obtained using the Gaussian entanglement Hamiltonian. To solve this problem, we propose that truncation of a Gaussian state should be performed simultaneously on all virtual bonds rather than on each bond in a consecutive manner. This strategy is not guaranteed to provide the most accurate MPS approximation for a given bond dimension, but our numerical experiments suggest that the results are satisfactory provided that $D$ is sufficientey large. For FGSs with relatively small entanglement (or with small system sizes), this method is already sufficient. However, for FGSs that require large system sizes to obtain a converged MPS local tensor, this method is not efficient enough. In Sec.~\ref{subsec:methodD} below, we describe an improved method for calculating the MPS local tensor.

As a further note, we mainly focus in this manuscript on Gutzwiller projected U(1)-FGSs as trial wave functions for spin systems, where a local fermion-number constraint restores the spin Hilbert space (e.g., singly occupied two-component fermions at each site representing spin-1/2). With a fixed number of local physical modes implemented by the Gutzwiller projection, virtual states at each sites also have a fixed particle number after Gutzwiller projection. This can obviate the need to compute unnecessary MPS local tensor entries.

When constructing fermionic tensor networks, minus signs in fermionic anticommutation relations should be handled carefully. For practical calculations, one needs to contract the virtual modes in Eq.~\eqref{eq:local-Am-2} with those in the virtual bond states to obtain an ordinary MPS, during which fermionic signs arise due to moving and contracting the virtual modes. However, these signs can be absorbed into the definition of the MPS local tensor (see Ref.~\cite{Kraus2010b}). Specifically, the sign factor
\begin{align}
(-1)^{\sum_{\beta} n_{r,\beta}\sum_{\gamma} n_{l,\gamma}+(\sum_{\beta} n_{r,\beta})^2} 
\end{align}
should be attached to each element $I_{\{n_{r,\beta}\},\{n_{l,\gamma}\}}$. Its value can be obtained by noting that the number of total virtual modes is $\sum_{\beta}n_{r,\beta}+\sum_{\gamma}n_{l,\gamma}$ and the total parity of the physical modes traversed by the virtual state is $(-1)^{\sum_{\beta}n_{r,\beta}}$. For the coefficients of $|S_{m,m+1}\rangle$, similar signs should also be attached as well. After taking these factors into account, the coefficients mentioned above simply serve as the local tensors.

\subsection{Mode decimation}
\label{subsec:methodD}

In Sec.~\ref{subsec:methodA}, \ref{subsec:methodB}, and \ref{subsec:methodC}, we have described a complete conversion algorithm, as we start from a U(1)-FGS $|\psi\rangle$ and arrive at an explicit form of the MPS local tensor.  There, the elements of the MPS local tensor are given by the Slater determinants of some sub-matrix of $S$ [Eqs.~\eqref{eq:diagonal} and \eqref{eq:occupied-d-mode}], so the computational complexity scales as $O(N^3)$, where $N$ is the system size. This scaling is polynomial but may still prevent us from simulating large systems. As reaching large system sizes is crucial for obtaining iMPS, the present subsection provides an improved method to compute the MPS local tensor. 

If we follow the SVD truncation scheme in Sec.~\ref{subsec:methodC} and restrict the number of virtual states in Eq.~\eqref{eq:local-Am-2} to $D$ (by choosing elements in $\{n_{l,\alpha}\}$ and $\{n_{r,\beta}\}$ with dominant entanglement support), some virtual modes ($b^{\dag}_{m,l,\alpha}$ and $b^{\dag}_{m,r,\beta}$) are occupied or unoccupied [see Eq.~\eqref{eq:S}; for these modes $\Lambda_{\mu}$ are close to $\pm 1$]. We estimate that only around $\mathrm{log}_2(D)$ modes are ``active'' (with $\Lambda_{\mu}$ not close to $\pm 1$) in these $D$ virtual states. With this observation in mind, we find that it is possible to define another local U(1)-FGS $|\tilde{A}^{[m]}\rangle$ which has less virtual modes (and thus a smaller correlation matrix $\tilde{\mathcal{C}}^{[m]}$) but gives exactly the same result when calculating the MPS local tensor with bond dimension $D$. As the size of $\tilde{\mathcal{C}}^{[m]}$ is of order $\mathrm{log}_2(D)$, the cost of generating a single MPS local tensor element can be reduced from $O(N^3)$ to $O((\mathrm{log}_2(D))^3)$. This is also more efficient than $O(N^2\mathrm{log}_2(D))$, where the latter requires a fast update strategy for Slater determinants (see Ref.~\cite{Petrica2021}).

In the following, we describe how to construct $|\tilde{A}^{[m]}\rangle$ by mode decimation and generate the MPS local tensor from it. Hereafter, we use a tilde hat for quantities where the mode decimation has been performed.

\begin{figure}
\centering
\includegraphics[width=1.0\linewidth]{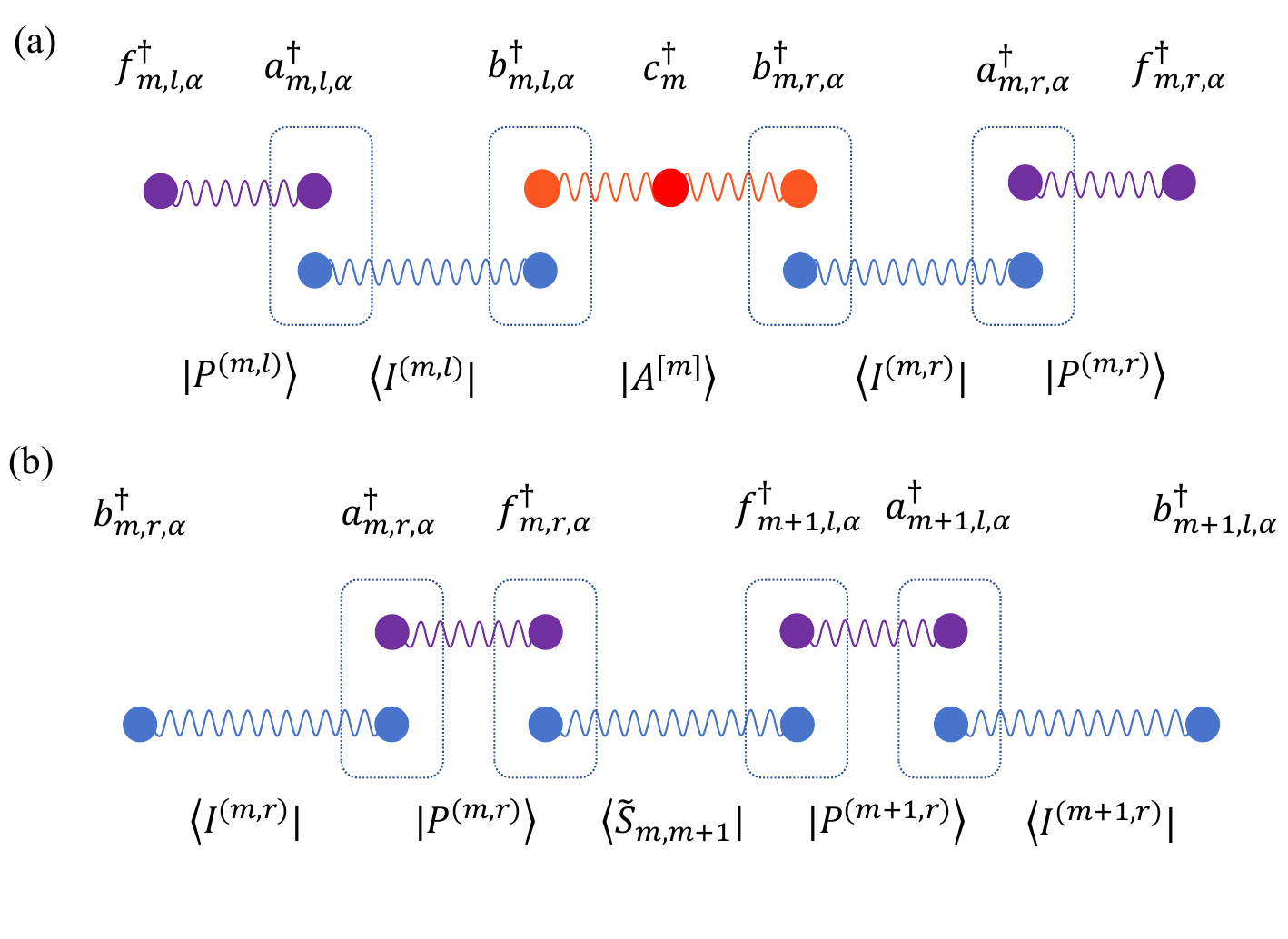}
\caption{Schematic of the Gaussian states (a) $|\tilde{A}^{[m]}\rangle$ and (b) $\langle S_{m,m+1}|$. The solid dots represent fermionic modes. The modes connected by a spiral line belong to the same Gaussian state. Different dots or modes in a single dashed rectangular are contracted.}
\label{fig:truncation}
\end{figure}

Due to the SVD truncation, some entangled modes in the Schmidt decomposition become occupied (unoccupied) modes.
To simplify the notation, we transfer this kind of entangled modes from set $\mathrm{E}$ to set $\mathrm{O}$ ($\mathrm{U}$) and denote the resulting set as $\tilde{\mathrm{E}}$, $\tilde{\mathrm{O}}$ and $\tilde{\mathrm{U}}$. Then, the local Gaussian state under mode decimation can be expressed as
\begin{align}
|\tilde{A}^{[m]}\rangle= (\langle I^{(m,l)}|\otimes \langle I^{(m,r)}|) (|P^{(m,l)}\rangle \otimes |A^{[m]}\rangle \otimes |P^{(m,r)}\rangle)
\label{eq:truncatedA}
\end{align}
with
\begin{align}
|P^{(m,l)}\rangle &=\prod_{\beta \in \tilde{\mathrm{U}}}a^{\dagger}_{m,l,\beta}\prod_{\alpha \in \tilde{\mathrm{E}}} (f^{\dagger}_{m,l,\alpha}-a^{\dagger}_{m,l,\alpha})|0\rangle, \nonumber \\
|I^{(m,l)}\rangle &=\prod_{\alpha=1}^{m-1} (a^{\dagger}_{m,l,\alpha}+b^{\dagger}_{m,l,\alpha}) |0\rangle, \nonumber \\
|I^{(m,r)}\rangle &=\prod_{\alpha=1}^{m} (b^{\dagger}_{m,r,\alpha}+a^{\dagger}_{m,r,\alpha}) |0\rangle, \nonumber \\
|P^{(m,r)}\rangle &=\prod_{\alpha \in \tilde{\mathrm{E}}}(a^{\dagger}_{m,r,\alpha}-f^{\dagger}_{m,r,\alpha}) \prod_{\beta \in \tilde{\mathrm{U}}}a^{\dagger}_{m,r,\beta} |0\rangle.  
\end{align}
Here we have introduced two new types of virtual modes, denoted by $f^{\dagger}$ and $a^{\dagger}$, their positions are depicted in Fig.~\ref{fig:truncation}(a). The projectors $P^{(m,r)}$ and $P^{(m,l)}$ can help fix or freeze modes belonging to $\tilde{\mathrm{U}}$ or $\tilde{\mathrm{O}}$. As $f^{\dagger}_{\mu} \in \tilde{\mathrm{E}}$, the number of $f$ modes can be smaller than that of $b$ modes, so $|\tilde{A}^{[m]}\rangle$ has less virtual modes.

To calculate tensor elements using $|\tilde{A}^{[m]}\rangle$, we should also perform mode decimation on the virtual states. For instance, if we start with the virtual state \{101100\} and find that the first and the last modes should be fixed, then the virtual state for $|\tilde{A}^{[m]}\rangle$ should be \{0110\}. Then $|P^{(m,r)}\rangle$ together with $\langle I^{(m,r)}|$ can be treated as a ``signal processor". If the input on the right side is \{0110\} (configuration of $f^{\dagger}_{m,r,\alpha}$), the output on the left side is \{101100\} (configuration of $b^{\dagger}_{m,r,\alpha}$). Following this procedure, one can easily verify that calculating the local tensor using $|\tilde{A}^{[m]}\rangle$ or $|A^{[m]}\rangle$ gives the same result. 

The block structure of $\tilde{\mathcal{C}}^{[m]}$ is much more complicated than that of $\mathcal{C}^{[m]}$ as some blocks no longer vanish due to mode decimation. To derive $\tilde{\mathcal{C}}^{[m]}$, we should first obtain the correlation matrices for those truncation projectors.
The correlation matrix of $|P^{(m,r)}\rangle$ can be easily calculated:
\begin{align}
\mathcal{C}^{(m,r)}=
\begin{pNiceMatrix}[first-row, first-col]
 & \scriptstyle{a}& \scriptstyle{f} \\[1mm]
\scriptstyle{a} & \mathcal{C}^{(m,r)}_{1,1} &  \mathcal{C}^{(m,r)}_{1,2}\\[2mm]
\scriptstyle{f} & [\mathcal{C}^{(m,r)}_{1,2}]^{\dagger} &  \mathcal{C}^{(m,r)}_{2,2}
\end{pNiceMatrix},
\end{align}
where
\begin{align}
\mathcal{C}^{(m,r)}_{2,2} &= 0,\\
[\mathcal{C}^{(m,r)}_{1,2}]_{\alpha,\beta} &=-\delta_{\alpha\beta},\\
[\mathcal{C}^{(m,r)}_{1,1}]_{\alpha,\beta}&=\delta_{\alpha\beta}
\left\{
\begin{aligned}
-1&,& \alpha \in \tilde{\mathrm{U}} \\
1&,& \alpha \in \tilde{\mathrm{O}}\\
0&,& \alpha \in \tilde{\mathrm{E}}
\end{aligned}\right. ,
\end{align}
and the correlation matrix of $|P^{(m,l)}\rangle$ reads
\begin{align}
\begin{split}
\mathcal{C}^{(m,l)}&=
\begin{pNiceMatrix}[first-row, first-col]
 & \scriptstyle{f}& \scriptstyle{a} \\[1mm]
\scriptstyle{f} & \mathcal{C}^{(m,l)}_{1,1} &  \mathcal{C}^{(m,l)}_{1,2}\\[2mm]
\scriptstyle{a} & [\mathcal{C}^{(m,l)}_{1,2}]^{\dagger} &  \mathcal{C}^{(m,l)}_{2,2} 
\end{pNiceMatrix},
\end{split}
\end{align}
where
\begin{align}
\mathcal{C}^{(m,l)}_{1,1} &= 0, \nonumber \\
[\mathcal{C}^{(m,l)}_{1,2}]_{\alpha,\beta} &=-\delta_{\alpha\beta}, \nonumber \\
[\mathcal{C}^{(m,l)}_{2,2}]_{\alpha,\beta}&=\delta_{\alpha\beta}
\left\{
\begin{aligned}
1&,& \alpha \in \tilde{\mathrm{U}} \\
-1&,& \alpha \in \tilde{\mathrm{O}}\\
0&,& \alpha \in \tilde{\mathrm{E}}
\end{aligned}\right. .
\end{align}
Combining Eq.~(A14) in Ref.~\cite{LiJW2023} and Eq.~\eqref{eq:truncatedA}, we obtain each block of the correlation matrix $\tilde{\mathcal{C}}^{[m]}$ [which also takes a $3\times3$ block structure as $\mathcal{C}^{[m]}$ in Eq.~\eqref{eq:A-CM}]:

\begin{widetext}
\begin{align}
\tilde{\mathcal{C}}^{[m]}_{1,1}&=\mathcal{C}^{(m,l)}_{1,2}\{\mathbbm{1}-\mathcal{C}^{[m]}_{1,3}\mathcal{C}^{(m,r)}_{1,1}[\mathcal{C}^{[m]}_{1,3}]^{\dagger}\mathcal{C}^{(m,l)}_{2,2}\}^{-1} \mathcal{C}^{[m]}_{1,3}\mathcal{C}^{(m,r)}_{1,1}[\mathcal{C}^{[m]}_{1,3}]^{\dagger} [\mathcal{C}^{(m,l)}_{1,2}]^{\dagger}, \nonumber \\
\tilde{\mathcal{C}}^{[m]}_{1,2}&=-\mathcal{C}^{(m,l)}_{1,2}\{\mathbbm{1}-\mathcal{C}^{[m]}_{1,3}\mathcal{C}^{(m,r)}_{1,1}[\mathcal{C}^{[m]}_{1,3}]^{\dagger}\mathcal{C}^{(m,l)}_{2,2}\}^{-1} \mathcal{C}^{[m]}_{1,3}\mathcal{C}^{(m,r)}_{1,1}[\mathcal{C}^{[m]}_{2,3}]^{\dagger}, \nonumber \\
\tilde{\mathcal{C}}^{[m]}_{1,3}&=\mathcal{C}^{(m,l)}_{1,2}\{\mathbbm{1}-\mathcal{C}^{[m]}_{1,3}\mathcal{C}^{(m,r)}_{1,1}[\mathcal{C}^{[m]}_{1,3}]^{\dagger}\mathcal{C}^{(m,l)}_{2,2}\}^{-1} \mathcal{C}^{[m]}_{1,3} \mathcal{C}^{(m,r)}_{1,2}, \nonumber \\
\tilde{\mathcal{C}}^{[m]}_{2,2}&=\mathcal{C}^{[m]}_{2,3} \mathcal{C}^{(m,r)}_{1,1}[\mathcal{C}^{[m]}_{2,3}]^{\dagger}+\mathcal{C}^{[m]}_{2,3} \mathcal{C}^{(m,r)}_{1,1} [\mathcal{C}^{[m]}_{1,3}]^{\dagger}\{\mathbbm{1}-\mathcal{C}^{(m,l)}_{2,2}\mathcal{C}^{[m]}_{1,3}\mathcal{C}^{(m,r)}_{1,1}[\mathcal{C}^{[m]}_{1,3}]^{\dagger}\}^{-1}\mathcal{C}^{(m,l)}_{2,2}\mathcal{C}^{[m]}_{1,3}\mathcal{C}^{(m,r)}_{1,1}[\mathcal{C}^{[m]}_{2,3}]^{\dagger}, \nonumber \\
\tilde{\mathcal{C}}^{[m]}_{2,3}&=-\mathcal{C}^{[m]}_{2,3} \mathcal{C}^{(m,r)}_{1,2}-\mathcal{C}^{[m]}_{2,3} \mathcal{C}^{(m,r)}_{1,1} [\mathcal{C}^{[m]}_{1,3}]^{\dagger}\{\mathbbm{1}-\mathcal{C}^{(m,l)}_{2,2}\mathcal{C}^{[m]}_{1,3}\mathcal{C}^{(m,r)}_{1,1}[\mathcal{C}^{[m]}_{1,3}]^{\dagger}\}^{-1}\mathcal{C}^{(m,l)}_{2,2}\mathcal{C}^{[m]}_{1,3}\mathcal{C}^{(m,r)}_{1,2}, \nonumber \\
\tilde{\mathcal{C}}^{[m]}_{3,3}&=-[\mathcal{C}^{(m,r)}_{1,2}]^{\dagger}[\mathcal{C}^{[m]}_{1,3}]^{\dagger}\{\mathbbm{1}-\mathcal{C}^{(m,l)}_{2,2}\mathcal{C}^{[m]}_{1,3}\mathcal{C}^{(m,r)}_{1,1}[\mathcal{C}^{[m]}_{1,3}]^{\dagger}\}^{-1}  \mathcal{C}^{(m,l)}_{2,2}\mathcal{C}^{[m]}_{1,3} \mathcal{C}^{(m,r)}_{1,2}.
\label{eq:truncateC}
\end{align} 
\end{widetext}

The mode decimation in the virtual bond states can be performed in a similar way. Here we only discuss $\langle \tilde{S}_{m,m+1}|$, and the calculation for $\langle \tilde{I}_{m,m+1}|$ is similar. To make sure that $\langle \tilde{S}_{m,m+1}|$ gives the same result as $\langle S_{m,m+1}|$ when generating the local matrix on the virtual bond, we require that $\langle S_{m,m+1}|=(\langle I^{(m,r)}| \otimes \langle \tilde{S}_{m,m+1} |\otimes \langle I^{(m+1,l)}|)\ (|P^{(m,r)}\rangle \otimes |P^{(m+1,l)}\rangle)$; see Fig.~\ref{fig:truncation}(b). It turns out that the virtual bond state after mode decimation is given by
\begin{align}
|\tilde{S}_{m,m+1}\rangle &=\prod_{\alpha \in \tilde{\mathrm{E}}} \left[ \sqrt{\frac{1-\Lambda_{\alpha}}{2}}f^{\dagger}_{m,r,\alpha}-\sqrt{\frac{1+\Lambda_{\alpha}}{2}}f^{\dagger}_{m+1,l,\alpha} \right] |0\rangle.
\end{align}

Before closing this subsection, we would like to emphasize that no error is generated during the mode decimation. The source of approximation is the SVD truncation discussed in Sec.~\ref{subsec:methodC} whereas mode decimation is designed to accelerate calculation. The key point is that $|\tilde{A}\rangle$ only provides us correct results for those elements that can survive in SVD truncation. If the bond dimension gets larger, the numbers of surviving elements and active modes also increase, and $|\tilde{A}\rangle$ should be modified accordingly.

\subsection{Filtration of anyon eigenbasis}
\label{subsec:methodE}

Many Gutzwiller projected FGSs in two dimensions have intrinsic topological order, and this is also so for the two examples which we will consider subsequently. Generally, the ground states for topologically ordered systems on a cylinder are the superposition of anyon eigenstates. In the language of MPS, the MPS representation of any superposition of anyon eigenstates is non-injective and the leading eigenvalues of its transfer matrix are degenerate. The same is true for systems with spontaneous symmetry breaking.

If the Gutzwiller projected FGSs have already been converted to iMPSs, we can filter out the degenerate ground states by using the fixed points of the iMPS transfer matrix. Let us denote the transfer matrix by $\mathbb{T}$. If the iMPS has a multi-site unit cell, $\mathbb{T}$ is the transfer matrix for the whole unit cell. Starting with some ``product state'' Ans\"atze $E_{\mathrm{trial}}$, we apply the transfer matrix on them and iterate many times to search for \emph{distinct} fixed points (as left eigenvectors of $\mathbb{T}$), $\lim_{n \rightarrow \infty} E_{\mathrm{trial}}\mathbb{T}^{n} \rightarrow E_{\zeta}$ with $\zeta = 1,\ldots,d_{\mathrm{deg}}$, where $d_{\mathrm{deg}}$ is the number of distinct fixed points (equivalently, the number of degenerate ground states). These fixed points can be normalized such that $E_{\zeta} E^{\dag}_{\zeta^{\prime}} = \delta_{\zeta,\zeta^{\prime}}$.

\begin{figure}
    \centering
    \includegraphics[width=1.0\linewidth]{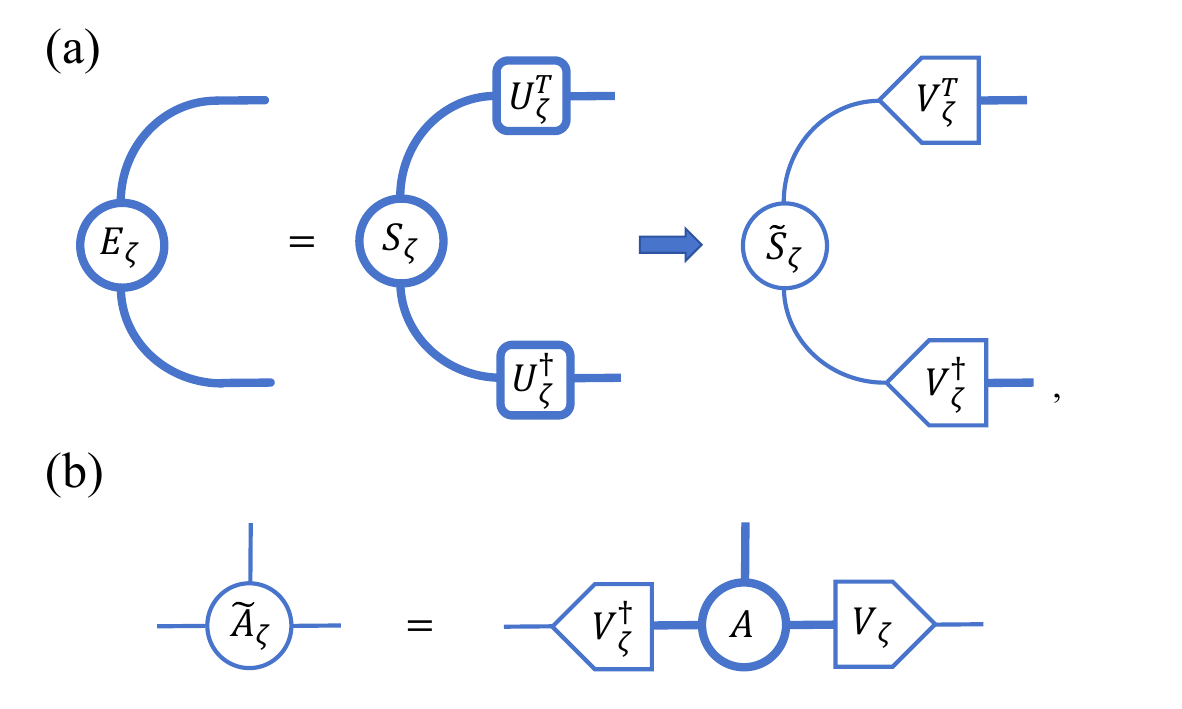}
    \caption{(a) Diagonalize the fixed points $E_{\zeta}$ and truncate the vanishing eigenvalues to obtain the isometries $V_{\zeta}$. (b) Use the isometries to obtain the iMPS local tensor for each degenerate ground state.}
    \label{fig:filtration}
\end{figure}

Once all distinct fixed points have been found, we use the following steps to obtain the iMPS local tensor for each ground state: (i) Diagonalize each fixed point $E_{\zeta}$, $E_{\zeta} = U_{\zeta} S_{\zeta} U^{\dagger}_{\zeta}$ with $S_{\zeta} \geq 0$. (ii) Truncate $S_{\zeta}$ (remove its zero eigenvalues) to obtain $E_{\zeta} = V_{\zeta} \tilde{S}_{\zeta} V^{\dagger}_{\zeta}$, where $\tilde{S}_{\zeta}>0$ encodes positive eigenvalues. $V_{\zeta}$ are isometries satisfying $V_{\zeta}V^{\dag}_{\zeta} = \mathbbm{1}$ and $V^{\dag}_{\zeta} V_{\zeta} = \mathcal{P}_{\zeta}$, where $\mathcal{P}_{\zeta}$ are orthogonal projectors satisfying a completeness relation $\sum_{\zeta=1}^{d_{\mathrm{deg}}} \mathcal{P}_{\zeta} = \mathbbm{1}$. (iii) Use the isometries to project onto the virtual space. If we denote the iMPS local matrix as $A$ (physical index omitted; it may correspond to a block of local tensors if one deals with a multi-site unit cell), the iMPS local tensor for each degenerate ground state is given by $\tilde{A}_{\zeta} = V^{\dag}_{\zeta} A V_{\zeta}$. The full procedure is depicted in Fig.~\ref{fig:filtration}.

\section{Results}
\label{sec:results}

This section presents our numerical results in two chiral spin liquid models. The first (second) one has the same topological order as the Laughlin (Moore-Read) state of bosons at filling factor $1/2$ ($1$)~\cite{Laughlin1983,Moore1991}. In both cases, the anyon eigenbasis is constructed and their entanglement spectra are computed~\cite{LiH2008}.  

\subsection{$S=1/2$ Abelian chiral spin liquid}
\label{subsec:resultA}
\begin{figure*}
    \centering
    \includegraphics[width=1.0\linewidth]{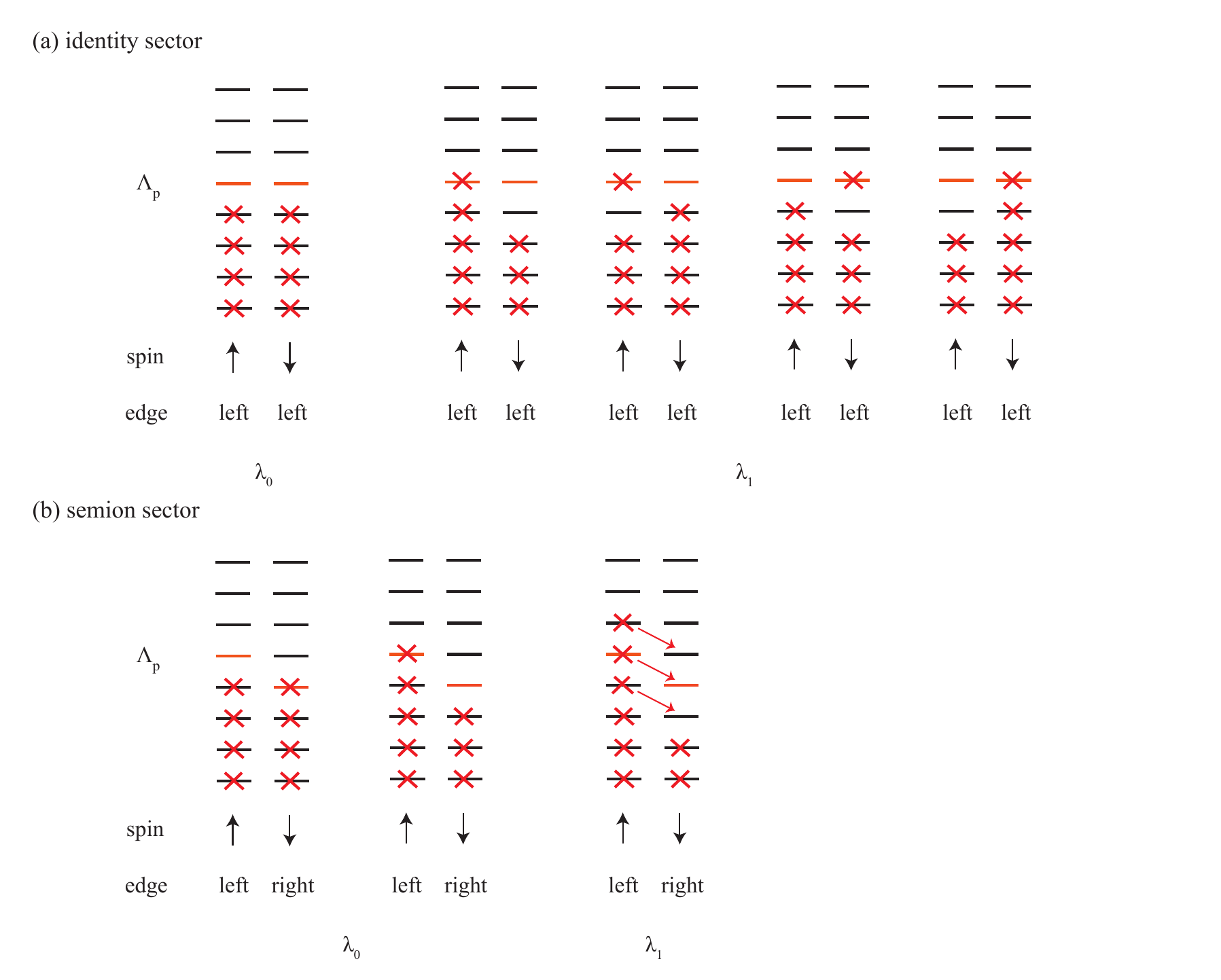}
    \caption{$\Lambda_p$ level for spin-up and spin-down orbitals in (a) the identity sector and (b) the semion sector of the spin-1/2 Laughlin chiral spin liquid. Only eight single-particle levels are plotted for depiction and those with $\Lambda_p=0$ are marked in orange. The Schmidt vectors with the largest and second largest singular value, denoted by $\lambda_0$ and $\lambda_1$, respectively, are given and the red crosses represent the occupied virtual mode. In (b), we use arrows to indicate the eight degenerate virtual states at $\lambda_1$ (only one of the two modes connected by an arrow is occupied in each virtual state, so we have $2^3$ virtual states). }
    \label{fig:SU2}
\end{figure*}
The first model is defined on the square lattice with each site occupied by one spin-1/2~\cite{ZhangY2011,TuHH2013a,WuYH2020}. There are $N_x$ ($N_y$) sites along the $x$ ($y$) direction. The spin-1/2 operators are expressed in terms of fermionic parton operators as
\begin{align}
S^{a}_{j} = \frac{1}{2} \sum_{\alpha,\beta=\uparrow,\downarrow} c^{\dag}_{j,\alpha} \sigma^{a}_{\alpha,\beta} c_{j,\beta},
\end{align}
where $j$ is the lattice site index, $\sigma^{a}$ ($a=x,y,z$) are the Pauli matrices, and $c_{j,\alpha}$ ($c^{\dag}_{j,\alpha}$) is the fermionic annihilation (creation) operator. A single-occupancy constraint $\sum_{\alpha=\uparrow,\downarrow} c^{\dag}_{j,\alpha} c_{j,\alpha} = 1$ ensures that the physical states at each site are spin-1/2 states, represented by $|\uparrow_j\rangle = c^{\dag}_{j,\uparrow}|\mathrm{vac}\rangle$ and $|\downarrow_j\rangle = c^{\dag}_{j,\downarrow}|\mathrm{vac}\rangle$, where $|\mathrm{vac}\rangle$ is the vacuum of fermionic partons.

\begin{figure*}[ht] 
\centering
\includegraphics[width=1.0\textwidth]{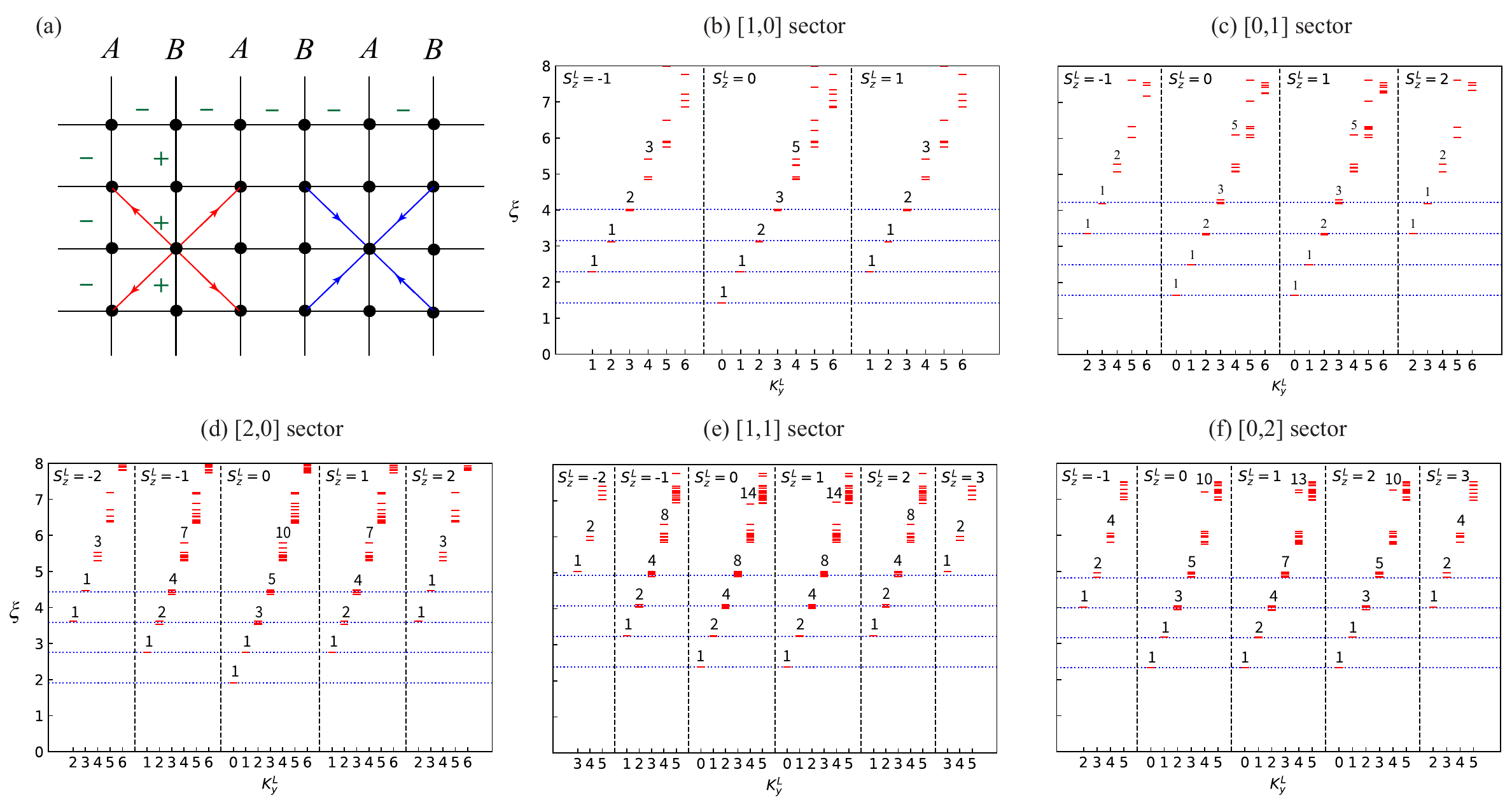}
\caption{(a) Schematics of signs of $t_{i,j}$ and $\Delta_{i,j}$ in parton Hamiltonian $H$. The signs of $t_{i,j}$ are indicated by the $\pm$ along the associated link. The signs of $\Delta_{i,j}$ are negative (positive) along (against) the arrows on the colored lines. Entanglement spectrum of (b) the identity sector and (c) the semion sector of the $\mathrm{SU}(2)_1$ chiral spin liquid. Entanglement spectrum of (d) the identity sector, (e) the Ising anyon sector, and (f) the fermion sector of the $\mathrm{SU}(2)_2$ chiral spin liquid.}
\label{fig:EE}
\end{figure*}

The hopping Hamiltonian for the fermionic partons is defined as~\cite{ZhangY2011}
\begin{align}
\label{eq:PartonHamt}
H = \sum_{\langle jk \rangle,\alpha} t_{jk}c^{\dagger}_{j,\alpha}c_{k,\alpha}+\sum_{\langle\langle jk \rangle\rangle,\alpha} i\Delta_{jk}c^{\dagger}_{j,\alpha}c_{k,\sigma} \, ,
\end{align}
where $\langle jk \rangle$ and $\langle\langle jk \rangle\rangle$ denote nearest and next-nearest neighbors, respectively. The hopping parameters have absolute values $|t_{jk}|=1.0$ and $|\Delta_{jk}|=0.5$ and their signs are displayed in Fig.~\ref{fig:EE}(a). The lower band with Chern number $1$ for each spin component is fully occupied. This Fermi sea is acted on by Gutzwiller projector $P_{\mathrm{G}}$ to generate the desired chiral spin liquid. 

To characterize the SU(2)$_1$ topological order of this model, it is convenient to employ cylinders that is open along the $x$ direction and periodic along the $y$ direction with boundary twist angle $\theta_{y}$. As already demonstrated in previous works, the anyon eigenbasis can be constructed easily when the parton Hamiltonian has exact zero modes localized at the two ends of the cylinders~\cite{TuHH2013a,WuYH2020}. If $N_y \in 4\mathbb{N}^{+}$ and $\theta_y=0$ or $N_{y} \in 4\mathbb{N}^{+}-2$ and $\theta_y=\pi$, there are four zero modes on either side of the cylinder (two for each spin component of the partons). The anyon eigenstate in the identity sector is obtained when the spin-up and spin-down zero modes on one end of the cylinder are occupied. The anyon eigenstate in the semion sector is obtained when the spin-up zero mode on one end and spin-down zero mode on the other end are occupied. 

One significant advance enabled by our method is that there is no need to manually select the edge modes. Instead, with the iMPS in hand, it is possible to filter out the anyon eigenbasis by analyzing the fixed points of the transfer matrix.
When $\theta_y \neq 0,\pi$, there is no exact zero mode. Tuning $\theta_y$ from $0$ to $\pi$, we numerically observed that only the identity sector is obtained. To get the semion sector, one needs to slightly couple the edge modes on each side. Two sectors get mixed in this case and the filtration scheme works. This procedure has been carried out in various systems and we focus on the case with $N_{x}=64$ and $N_{y}=10$. The lattice is mapped to a chain with a snake shape to accommodate the MPS. When the system is sufficiently long along the $x$ direction, the MPS tensors at the center can be taken out to serve as unit cells of iMPS. There are $20$ sites in each unit cell because the hopping Hamiltonian has period $2$ along the $x$ direction. We have constructed the iMPS representation with bond dimension $D=800$ for both anyon eigenstates. To compute the entanglement spectrum, the system is divided into left and right halves so the good quantum numbers are $K^{L}_{y}$ (the total momentum along $y$) and $S^{L}_{z}$ (the azimuthal spin). The edge physics of the $1/2$ Laughlin state is captured by the chiral SU(2)$_{1}$ Wess-Zumino-Witten (WZW) model. This is a chiral conformal field theory (CFT) with two primary fields corresponding to affine Dynkin labels $[0,1]$ and $[1,0]$~\cite{Francesco-Book}. The energy spectrum of this model is well known. As shown in Figs.~\ref{fig:EE}(b) and (c), the counting of entanglement levels agrees with the CFT predictions.

It is helpful to inspect the spin symmetry of the anyon eigenstates. Both the parton Hamiltonian and Gutzwiller projection respect SU(2) spin rotation symmetry. In the Schmidt decomposition of an SU(2)-symmetric MPS, the Schmidt vectors associated with the same virtual block must have the same singular values. For the identity sector, this symmetry can be preserved by the anyon eigenstate as shown in Fig.~\ref{fig:SU2}(a). The largest singular value $\lambda_0$ is unique when a cut is made at the center, so it belongs to the spin-0 sector. The second-largest singular value $\lambda_1$ is four-fold degenerate, which means that the associated Schmidt vectors arise from tensor product of two spin-$1/2$ and their virtual block is $\frac{1}{2}\otimes\frac{1}{2} = 0 {\oplus} 1$. The other virtual blocks can be understood similarly. On the contrary, the SU($2$) symmetry is broken to U($1$) in the semion sector. This is already evident in Fig.~\ref{fig:EE}(c) as the entanglement levels with $S^{L}_{z}=\pm 1$ are not identical. If we construct a spin-singlet ansatz for the semion state, the SU(2) symmetry can be restored~\cite{WuYH2020}. The largest singular value $\lambda_{0}$ is two-fold degenerate so its virtual block has spin-$1/2$. The second largest singular value $\lambda_{1}$ corresponds to the virtual block $\frac{1}{2}\otimes\frac{1}{2}\otimes\frac{1}{2}=\frac{1}{2}\oplus \frac{1}{2} \oplus \frac{3}{2}$.

\subsection{$S=1$ non-Abelian chiral spin liquid}
\label{subsec:resultB}

The (bosonic) Moore-Read state is a canonical example of non-Abelian topological order associated with the chiral SU(2)$_{2}$ WZW model~\cite{Moore1991}. Its three anyon eigenstates are labeled as [2,0] (identity sector), [1,1] (Ising anyon sector), and [0,2] (fermion sector). In the quantum Hall contexts, the bosonic Moore-Read state can be constructed using two copies of the bosonic Laughlin state with proper symmetrization. It has been proposed that the same topological order can be realized in spin-1 models~\cite{Greiter2009,Greiter2014,Lecheminant2017,LiuZX2018,ChenJY2018,ZhangHC2021,Jaworowski2022,HuangYX2022} as well as the Yao-Lee model~\cite{YaoH2011}.

We use the fermionic parton approach to construct a spin-1 chiral spin liquid state which has the same topological order as the bosonic Moore-Read state. To this end, we adopt a two-orbital construction~\cite{Shastry1992} and write the spin-1 operators as
\begin{align}
S^a_j=\frac{1}{2}\sum_{\tau=1,2}\sum_{\alpha,\beta=\uparrow,\downarrow}c^{\dagger}_{j,\tau,\alpha} \sigma^{a}_{\alpha,\beta} c_{j,\tau,\beta},
\end{align}
where $\tau=1,2$ stands for two different orbitals at the same site. The local constraint for restoring the spin-1 Hilbert space is imposed by requiring $\vec{S}^2_j=2$, so that each orbital must be occupied by exactly one fermionic parton (carrying spin-1/2) and the two fermionic partons must form a spin-1 triplet. The three spin-1 states are expressed in terms of fermionic partons as follows: 
\begin{align}
|1_j\rangle &= c^{\dagger}_{j,1,\uparrow} c^{\dagger}_{j,2,\uparrow}|\mathrm{vac}\rangle, \nonumber \\
|-1_j\rangle &=c^{\dagger}_{j,1,\downarrow} c^{\dagger}_{j,2,\downarrow}|\mathrm{vac}\rangle, \nonumber \\
|\ 0_j\rangle &=\frac{1}{\sqrt{2}} (c^{\dagger}_{j,1,\downarrow}c^{\dagger}_{j,2,\uparrow}+c^{\dagger}_{j,1,\uparrow}c^{\dagger}_{j,2,\downarrow})|\mathrm{vac}\rangle.
\label{eq:spin-1-states}
\end{align}
Accordingly, the Gutzwiller projector $P_{\mathrm{G}}$ is defined to remove states other than those in Eq.~\eqref{eq:spin-1-states}.

The parton Hamiltonian of this system is identical to Eq.~\eqref{eq:PartonHamt}, except that the fermionic partons have an additional orbital index. In the ground state, the lower band with Chern number 1 for each spin and orbital component is fully occupied. The Gutzwiller projector is applied on top of the parton Fermi seas to generate a spin-1 wave function which is expected to exhibit the SU(2)$_2$ chiral topological order. However, unlike the SU(2)$_1$ case in Sec.~\ref{subsec:methodA}, it is a prior unclear how to define the anyon eigenbasis in the current spin-1 model.

Using our filtration scheme, three anyon eigenstates are constructed using bond dimension $D=1600$ and their entanglement spectra are displayed in Fig.~\ref{fig:EE}(d)--(f). The counting of entanglement levels agrees with theoretical predictions based on the chiral SU(2)$_{2}$ WZW model. This success is a highly nontrivial corroboration of our method. If the parton hopping Hamiltonian indeed has exact zero modes, we can write down the anyon eigenbasis explicitly. To begin with, all parton modes with negative energy are occupied to generate the state $|\Phi\rangle$.
The zero modes are created by the operators $d^{\dagger}_{s,\tau,\alpha}$, where $s=L,R$ indicates the left or right edge of the cylinder, $\tau=1,2$ stands for the two orbitals, and $\alpha=\uparrow,\downarrow$ represents spin-up or spin-down partons. The anyon eigenbasis is given by
\begin{align}
|\psi_{[2,0]}\rangle &= P_{\mathrm{G}}  \, d^{\dagger}_{L,1,\uparrow} d^{\dagger}_{L,1,\downarrow} d^{\dagger}_{L,2,\uparrow} d^{\dagger}_{L,2,\downarrow} |\Phi\rangle, \nonumber \\
|\psi_{[1,1]}\rangle &= P_{\mathrm{G}} \, d^{\dagger}_{L,1,\uparrow} d^{\dagger}_{L,1,\downarrow} d^{\dagger}_{L,2,\uparrow} d^{\dagger}_{R,2,\downarrow} |\Phi\rangle,  \nonumber \\
|\psi_{[0,2]}\rangle &= P_{\mathrm{G}} \, d^{\dagger}_{L,1,\uparrow} d^{\dagger}_{R,1,\downarrow} d^{\dagger}_{L,2,\uparrow} d^{\dagger}_{R,2\downarrow} |\Phi\rangle \, .
\label{eq:psi}
\end{align}
This construction is not unique. For instance, the state in the identity sector, $\psi_{[2,0]}$, can also be expressed as $P_{\mathrm{G}} \, d^{\dagger}_{L,1,\uparrow} d^{\dagger}_{L,1,\downarrow} d^{\dagger}_{R,2,\uparrow} d^{\dagger}_{R,2,\downarrow} |\Phi\rangle$.

The results in Eq.~\eqref{eq:psi} reveal how to relate the anyon eigenbasis of the bosonic Moore-Read states to that of the bosonic Laughlin states. If we interpret the two different orbitals as different ``copies'', the first line of Eq.~\eqref{eq:psi} shows that the identity sector of the Moore-Read state is obtained using two copies of the Laughlin state in the identity sector. If one (both) copy of the Laughlin state is replaced by its semion sector, we would obtain the Ising anyon (fermion) sector of the Moore-Read state. This idea may also be utilized to generate other SU(2)$_{k}$ non-Abelian chiral spin liquid ($k>2$). In particular, the $k=3$ case possesses Fibonacci anyons but there are not many studies in the spin liquid context~\cite{LuoWW2023}. 

\section{Summary and discussions} 
\label{sec:summary}

In summary, we have demonstrated that the (infinite) MPS representation of a particle-number-conserving FGS can be constructed efficiently using our algorithm. Thanks to the extensive usage of correlation matrices in the present work, computational cost is significantly reduced compared to previous studies so much larger system sizes can be reached. For a given MPS that describes a topologically ordered system, the anyon eigenbasis can be filtered out in a straightforward manner by analyzing the fixed points of the transfer matrix. This helps to eliminate the laborious process of adjusting the boundary conditions to search for proper edge mode occupations. The algorithm is applied to study two chiral spin liquids and the entanglement spectra of their anyon eigenbasis agree with the CFT predictions.

It would be very interesting to further test our algorithm. If the iMPSs on the cylinder are twisted in certain ways, the modular matrix could be computed to provide a quite comprehensive characterization~\cite{Cincio2013,TuHH2013a,Zaletel2013}. Quantum phase transitions of topological states are routinely investigated in the parton framework and numerical verification of many theoretical predictions could be facilitated by our method. When a microscopic model is studied using DMRG, a wisely chosen initial state may substantially speedup the calculation. For a 2D system with topological order, feeding the program with suitable parton wave functions rather than random states could help us to find the anyon eigenbasis more easily. A natural extension is to study the cases without U(1) symmetry associated with particle number conservation. This is important for studying parton wave functions with paired fermions. The correlation matrix formalism is still valid in such cases, but one important difference is that matrix elements now include anomalous correlators $\langle c_i c_j \rangle$ and $\langle c^{\dag}_i c^{\dag}_j \rangle$~\cite{JinHK2022a}. It is quite likely that only minimal modifications are needed to adapt our algorithm to these systems.

\textit{Note added}. In finalizing this manuscript, we became aware of the work \cite{LiKL2025}, which also discussed how to construct fermionic Gaussian MPSs from correlation matrices with mode truncation.

\section*{Acknowledgments}

T.L. thanks Tong-Zhou Zhao for helpful discussions. H.H.T. is grateful to Jan von Delft, Hui-Ke Jin, Jheng-Wei Li, Rong-Yang Sun, Lei Wang, Qi Yang, Yi Zhou for stimulating discussions and fruitful collaborations on related topics. T.L. and T.X. are supported by the National Natural Science Foundation of China under Grant No. 12488201. Y.H.W. is supported by the National Natural Science Foundation of China under Grant No. 12174130.

\input{refs.bbl}

\end{document}

%% file: refs.bbl
%